\documentclass[12pt]{iopart}

\usepackage{iopams}
\usepackage{graphicx}
\usepackage{caption}
\usepackage{cite}
\usepackage{color}

\begin{document}

\title[Dry transfer method for suspended graphene on lift-off-resist]{Dry transfer method for suspended graphene on lift-off-resist: simple ballistic devices with Fabry-P\'{e}rot interference
}

\author{Ying Liu\textsuperscript{1,2}, T. S. Abhilash\textsuperscript{2}, Antti Laitinen\textsuperscript{2}, Zhenbing Tan\textsuperscript{2,3}, Guan-jun Liu\textsuperscript{1}, Pertti Hakonen\textsuperscript{2}}

\address{\textsuperscript{1} Science and Technology on Integrated Logistics Support Laboratory, National University of Defense Technology, Changsha, 410073, P. R. China}
\address{\textsuperscript{2} Low Temperature Laboratory, Department of Applied Physics, Aalto University, Espoo, 02150, Finland}
\address{\textsuperscript{3} Shenzhen Institute for Quantum Science and Engineering, and Department of Physics, Southern University of Science and Technology, Shenzhen 518055, China}
\ead{gjliu342@126.com, pertti.hakonen@aalto.fi}
\vspace{5pt}
\begin{indented}
\item[]December 2018
\end{indented}

\begin{abstract}
We demonstrate a fabrication scheme for clean suspended structures using chemical-vapor-deposition-grown graphene and a dry transfer method on lift-off-resist-coated substrates to facilitate suspended graphene nanoelectronic devices for technology applications. It encompasses the demands for scalable fabrication as well as for ultra-fast response due to weak coupling to environment. The fabricated devices exhibited initially a weak field-effect response with substantial positive ($p$) doping which transformed into weak negative ($n$) doping upon current annealing at the temperature of 4 Kelvin. With increased annealing current, $n$-doping gradually decreased while the Dirac peak position approached zero in gate voltage. An ultra-low residual charge density of ${9\times10^8 \mathrm{ \ cm^{-2}}}$ and a mobility of ${1.9 \times 10^5 \mathrm{\ cm^2/Vs}}$ were observed. Our samples display clear Fabry-P\'{e}rot (FP) conductance oscillation which indicates ballistic electron transport. The spacing of the FP oscillations are found to depend on the charge density in a manner that agrees with theoretical modeling based on Klein tunneling of Dirac particles. The ultra-low residual charge, the FP oscillations with density dependent period, and the high mobility prove excellent quality of our suspended graphene devices. Owing to its simplicity, scalability and robustness, this fabrication scheme enhances possibilities for production of suspended, high-quality, two-dimensional-material structures for novel electronic applications.
\end{abstract}

%
\vspace{2pc}

%
\submitto{\NT}
%
\maketitle
%
%

\section{Introduction}
\label{intro}

Suspended graphene devices with attractive mechanical and electronic properties have been utilized during the past few years as a playground for fundamental physics and as a platform for novel electronic applications\cite{Ferrari2015}. Recent fractional quantum Hall effect experiments \cite{bolotin2009observation,kumar2018unconventional}, electron-optic\cite{Rickhas2015}, electron-phonon interaction measurements \cite{rickhaus2013ballistic,laitinen2014electron} and several ballistic transport studies \cite{Bolotin2008Temperature,du2008approaching,oksanen2014single,kumaravadivel2016signatures} have employed high-quality suspended graphene samples. These devices  act also as nanoelectromechanical systems (NEMS) \cite{J2007Electromechanical}, which facilitate e.g. tunable mechanical resonators \cite{chen2013graphene,de2016tunable}, radio frequency components \cite{song2011stamp}, as well as ultra-sensitive sensors for mass and force \cite{eichler2011nonlinear,weber2016force}.

The conventional fabrication method for making suspended graphene devices \cite{du2008approaching,bolotin2008ultrahigh} involves exfoliation of graphene onto a Si/SiO$_2$ substrate, followed by patterning of the contacts. Subsequent wet etching  of SiO$_2$ in buffered hydrofluoric acid (BHF) is used to release and suspend the graphene structure. After isotropic SiO$_2$ etching, the graphene device is prone to collapse, which reduces yield of the fabrication process. Furthermore, reactivity of several metals with BHF limits material choices for the electrical contacts to graphene; even with less reactive metals, BHF etch may compromise the overall electrical characteristics of the graphene device by degrading the contact performance.

Several works \cite{rickhaus2013ballistic,tombros2011large,Grushina2013A} have demonstrated fabrication of suspended graphene devices using polydimethylglutarimide (PMGI) based Lift-Off-Resist (LOR) as a sacrificial layer instead of the conventional SiO$_2$. This method provides a large yield, and it is also compatible with commonly used contact materials for graphene. In this work, we too employed LOR as a supporting layer for CVD-grown (Chemical Vapor Deposition) graphene which was deposited using an optimized dry transfer method for transferring isolated flakes of graphene. Comprehensive measurements on the effect of current annealing on the device quality were carried out. Electrical transport at sub-Kelvin temperatures was performed in order to verify the high quality of our suspended CVD-graphene devices.

\begin{figure}
  \centering
  \includegraphics[width=0.80\textwidth]{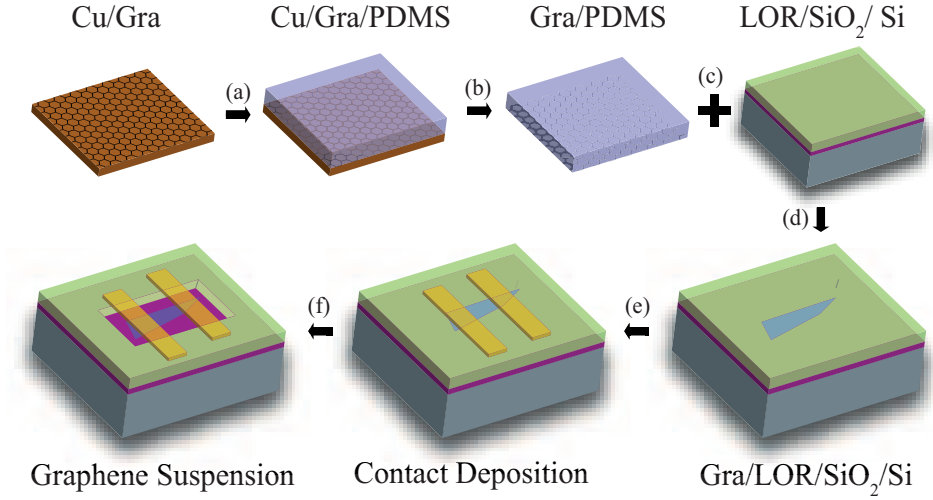}\\
  \caption{Schematic illustration of our CVD graphene (Gra) transfer and fabrication process: (a) adhesion of Cu/Gra to PDMS, (b) etching of the Cu, (c) deposition of LOR on to Si/SiO$_2$ substrate, (d) attachment of graphene/PDMS to LOR/SiO$_2$/Si, (e) metal contact deposition, (f) suspension of graphene.}\label{fig:1}
\end{figure}

\section{Device Fabrication Process}
\label{Fabrication}

The fabrication process flow is schematically depicted in figure \ref{fig:1}. First, a polydimethylsiloxane (PDMS) stamp was prepared and the graphene on copper foil was placed on the PDMS stamp with the graphene face down. The copper foil was gently pressed onto  the stamp in order to make graphene to adhere well to PDMS. The Cu/graphene/PDMS stamp was placed in Cu etchant (1M FeCl$_{3}$) to remove the copper, followed by repeated
rinsing in deionized water. To further remove etch residues from the graphene surface, the graphene-PDMS stamp was cleaned using the modified RCA method \cite{liang2011toward}. The substrate for graphene transfer was prepared by spin coating two layers of LOR-3A resist (PMGI based organic polymer from MicroChem) on $p^{++}$ silicon substrate capped with 285-nm-thick SiO$_2$. Each LOR layer resulted in a thickness of about 250 nm. The cleaned Graphene-PDMS stamp and the LOR coated substrate were then mounted onto a home-made dry transfer setup.

\begin{figure}
  \centering
  \includegraphics[width=0.80\textwidth]{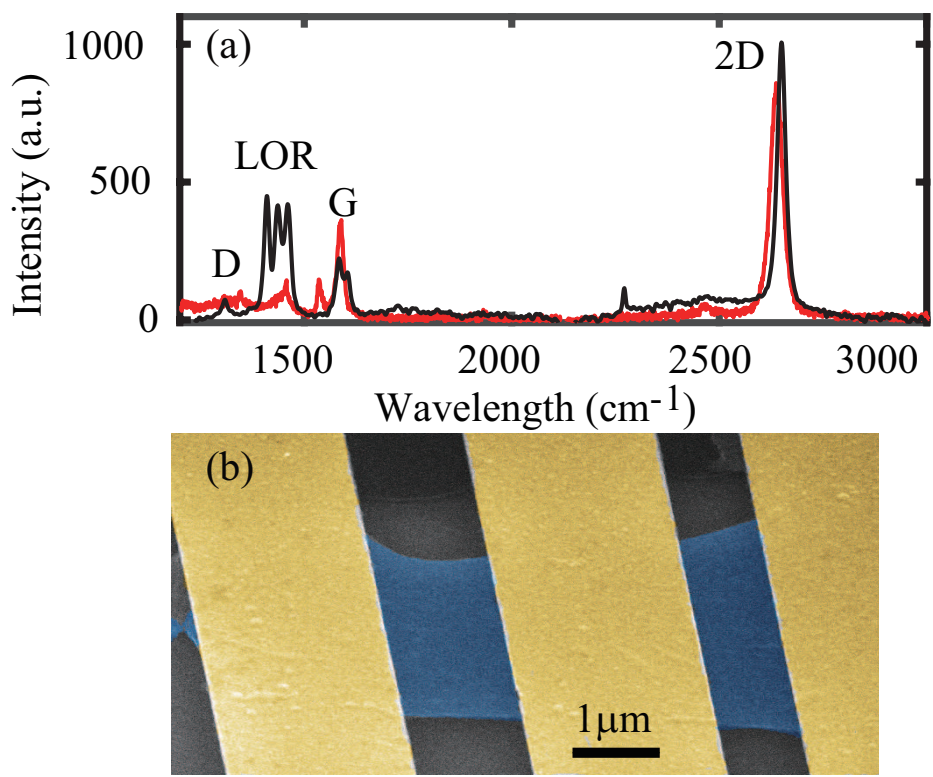}\\
  \caption{(a) Raman spectra of monolayer CVD graphene: the black indicates spectra of graphene transferred on to LOR, and the red indicates spectra of graphene on SiO$_2$/Si for comparison. (b) False color SEM image of fabricated suspended graphene device: metal contacts (yellow), graphene (blue).}\label{fig:2}
\end{figure}

The stamp was next aligned and brought into contact with the substrate, after which the stamp was peeled off of the substrate. When the stamp was peeled off slowly, uninterrupted graphene transfer could be achieved leading to the graphene covering the entire substrate. We note that this could be useful for larger scale fabrication of suspended graphene device arrays by introducing additional patterning and etching steps. On the other hand, when the stamp was immediately peeled off of the substrate with a high speed, then only small cracked flakes from continuous graphene film transferred due to incomplete adhesion to the substrate. Since these well-isolated graphene flakes could be immediately used for making suspended devices of size that can survive the suspension process without extra patterning and etching steps, we focus on them in the following experiments and analysis. Monolayer graphene flakes with suitable sizes were located using an optical microscope, followed by micro-Raman measurements as illustrated in figure \ref{fig:2}a. Signatures of graphene's characteristic Raman peaks labeled as D, 2D and G were observed, together with a LOR peak between the D and G peaks of graphene. The CVD graphene used here was foremost made of monolayer, but small patches of bi-layer were present. The bilayer regions resulted in darker optical contrast and the difference between mono and bi-layer could be easily distingushed using optical microscopy.

Two layers of polymethylmethacrylate (PMMA) (top most layer of 950K/A3, bottom layer of 50K/A11) was spin coated on the graphene/LOR chip for standard electron beam lithography. The lithography was operated under $20 \ \mathrm{keV}$ accelerating voltage. An electron beam dose of  $120 \ \mathrm{\mu C/cm^2}$ was employed to pattern the electrical leads. To enhance the selectivity of the development of the PMMA resist on top of LOR, the samples were developed in xylene instead of the regular developer methylisobutylketone (MIBK). The electrical leads were made by evaporating a Cr/Au bilayer with thickness 5/65 nm, respectively. The lift off process after evaporation was performed at 80 ${^\circ \mathrm C}$ in xylene. To release and suspend graphene, the LOR underneath the graphene and the metal contact regions was exposed to electron beam irradiation with a dose of $560 \ \mathrm{\mu C/cm^2}$, which allowed removal of the exposed LOR by immersing the sample in ethyl lactate. After rinsing in hexane, the sample was finally taken out from the low-surface-tension solvent and blown dry gently using N$_2$ gas. Figure \ref{fig:2}b, shows a false color Scanning Electron Microscope (SEM) image of one of the fabricated devices.

\section{Results and discussion}
\label{Results}

\begin{figure}
  \centering
  \includegraphics[width=0.80\textwidth]{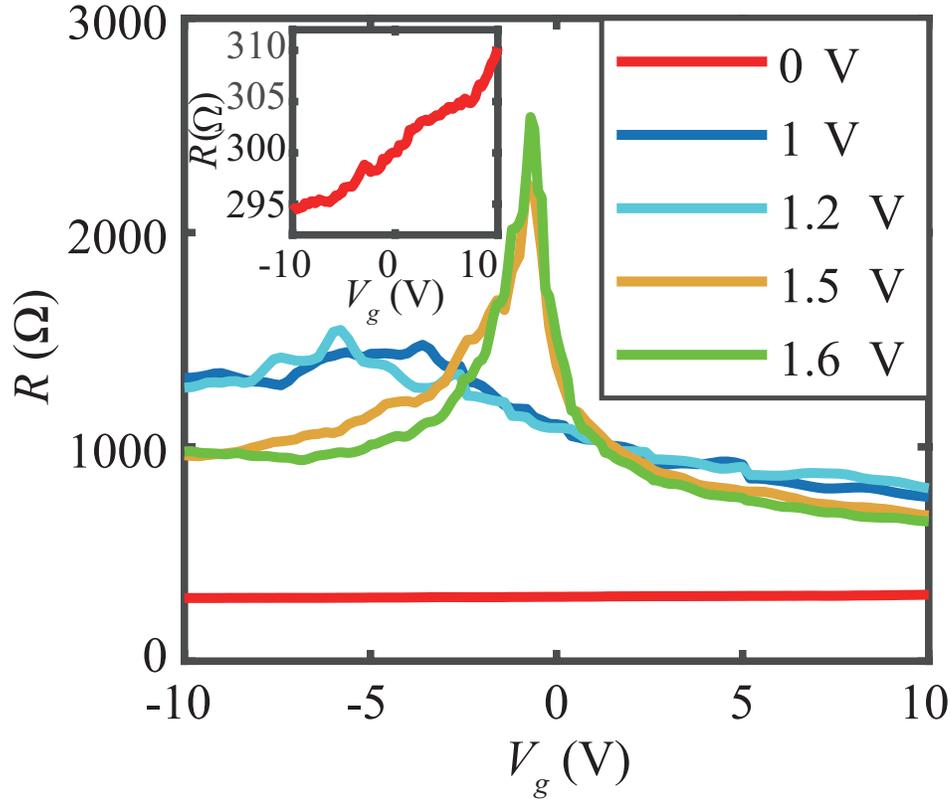}\\
  \caption{Resistance \textit{R} vs. gate voltage $V_g$ measured after \mbox{different} annealing voltage values indicated in the right inset. The left inset illustrates a magnified view of the gate sweep of the sample resistance before annealing (0 V).}\label{fig:3}
\end{figure}

The electronic transport measurements were conducted in a Bluefors LD250 dilution refrigerator. The devices were electrically characterized by measuring source-drain resistance as a function of back-gate bias with a fixed source-drain voltage. The gating curve of the as-fabricated device is shown as the red resistance trace in figure \ref{fig:3}; the inset displays a blow-up of the as-fabricated behavior. The absence of the Dirac peak and just a weak field effect  (see the inset) indicate strong positive doping ($p$-doping). Similar findings have been obtained in several other experiments \cite{tombros2011large,tombros2011quantized,pirkle2011effect,suk2013enhancement,kumar2013influence,kumar2014carrier,yang2012influence,romero2008n}, and they have  usually been attributed to  PMMA residues from the fabrication processes \cite{suk2013enhancement,kumar2013influence,kumar2014carrier} or ambient adsorbates like O$_2$, H$_2$O \cite{yang2012influence,romero2008n}. In our work, low-temperature current annealing was performed to remove contaminants and to improve  surface quality of graphene as described in reference \cite{kumar2018unconventional}. We measured the  resistance of the device as a function of the applied gate voltage $V_g$ after each annealing cycle. Upon current annealing, we see that the position of the Dirac point shifts to negative $V_g$ indicating $n$-doping, which is against the expected $p$-doping of graphene by PMMA residues. If PMMA as such were the $p$-dopant, then the device should consistently display $p$-doping until PMMA residues are completely removed. Intermediate-stage $n$-doping during annealing has also previously been observed in similar structures \cite{mizuno2013ballistic,grushina2015quantum,chen2015all}.

Instead of acting as an important $p$-doping source, PMMA residue is expected to play a neutral, benign role \cite{gammelgaard2014graphene,koo2015role} in establishing electrical properties of graphene. However, a reasonable explanation for the observed $n$-doping could be hydrogenation of graphene by H$_2^+$ and low-molecular-weight hydrocarbon radicals released from the PMMA decomposition \cite{ryu2008reversible,ennis2010mechanistical,kim2012n}.
These light and H-rich radicals released by resist induce the hydrogenation of graphene, which results in the $n$-doping at intermediate stages of the current annealing.

In order to understand the possible origin of H-rich radicals, let us first discuss the doping in samples on SiO$_2$ substrates. As is well known, pristine graphene surface is hydrophobic \cite{leenaerts2009water} and can hardly adsorb O$_2$ or H$_2$O molecules \cite{ryu2010atmospheric,xu2012investigating}. However, $p$-doping can be created by O$_2$ and H$_2$O that are adsorbed between graphene and SiO$_2$ interface, and are bonded by hydrogen bond with Si-OH on the surface of SiO$_2$ in on-substrate graphene devices.  This picture of doping has been verified by several experiments \cite{pirkle2011effect,ryu2010atmospheric,xu2012investigating,pinto2014electronic}.

In suspended graphene devices, we argue that the \mbox{PMMA} residue forms - before it is fully removed - a layer which plays a role similar to the SiO$_2$ for the adsorption of O$_2$, H$_2$O molecules. The initial $p$-doping observed in figure \ref{fig:3} would then originate from the dissolution of O$_2$, H$_2$O in the PMMA residue (or PMMA/graphene interface) rather than direct adsorption on the graphene surface. During current annealing, the graphene sheet heats up, which leads to evaporation of O$_2$ and H$_2$O molecules as well as decomposition of PMMA residues with the accompanying creation of H$_2^+$ and low-molecular-weight hydrocarbon radicals.
As a result, the $p$-doping in graphene gradually weakens and $n$-doping emerges.
As the annealing process continues, the graphene sheet recovers its nearly pristine, undoped characteristics, which is implied by the position of the Dirac peak at $V_g = -0.65$ V (figure \ref{fig:3}).

\begin{figure}
  \centering
  \includegraphics[width=0.80\textwidth]{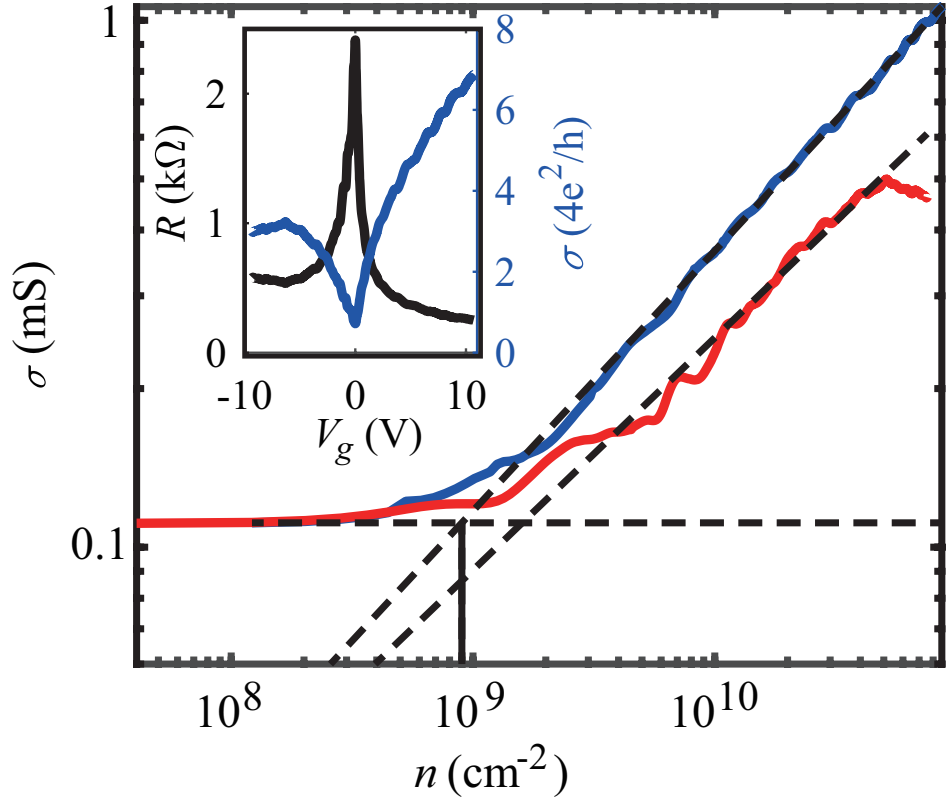}\\
  \caption{Conductivity $\sigma$ vs. charge carrier density $n$. The red and blue curves display the hole and electron transport data, respectively. Both data set have been fitted using $\sigma \propto n^{\gamma}$, and the results with $\gamma= 0.5$ and $\gamma= 0.44$ are shown by the dashed lines. The horizontal dashed line denotes the minimum level of conductivity. The vertical black line indicates the cross-over point between residual $n$ and $n^{\gamma}$ behavior. The inset displays $R_{reduce}$ and the corresponding $\sigma$ as a function of $V_g$ after subtracting the contact resistance.}\label{fig:4}
\end{figure}

\begin{figure}
  \centering
  \includegraphics[width=0.80\textwidth]{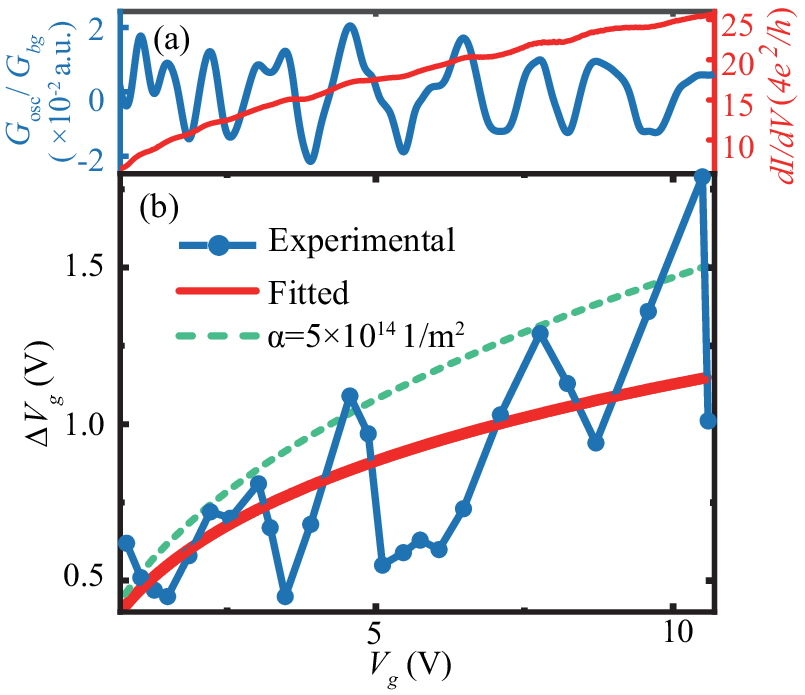}\\
  \caption{Fabry-P\'{e}rot (FP) oscillation and its theoretical fitting. (a) Differential conductance $dI/dV$ on the $n$-doping side vs. $V_g$ (red) and the relative oscillation $G_{osc}/G_{bg}$ (blue) for the extraction of the FP oscillation period. (b) Fabry-P\'{e}rot oscillation period  $\Delta V_g$ as a function of $V_g$ (blue); the red curve indicates a theoretical fitting based on the trapezoidal barrier model of reference \cite{Sonin2009Effect,Laitinen2015Klein}; the dashed green curve displays the theoretical calculation of equation \ref{period} with $a=5\times 10^{14}\ \mathrm{1/m^2}$ based on reference \cite{Laitinen2015Klein}.}\label{fig:5}
\end{figure}

The ideal conductivity $\sigma_{ideal}$ of a graphene device arises from mode-dependent,  transmission probabilities of massless Dirac fermions \cite{katsnelson2006zitterbewegung,tworzydlo2006sub}, which leads to a minimum $\sigma_0=\frac{4}{\pi}\frac{e^2}{h}$. In order to compare our results with theory, we subtract a contact resistance of 560 $\mathrm{\Omega \cdot \mu m}$ which guarantees a correct asymptotic behavior of $\sigma(V_g)$ at large $V_g$ of our electron side data. The reduced device resistance $R_{reduce}$ and the corresponding conductivity $\sigma$ as a function of $V_g$ are shown by the black  and blue curves in the inset of figure \ref{fig:4}, respectively. The minimum conductivity of our sample amounts to $\sigma_{min} = 0.72G_0$, where $G_0=(4e^2)/h$. This result is close to other experimental results on high-quality graphene (see e.g. reference \cite{du2008approaching}). For electron mobility we obtain  $\mu = 1.9 \times 10^5$ $\mathrm{cm^2/Vs}$ from the Drude model ($\sigma=ne\mu$) applied at carrier density of $n=4.3\times10^{9}\ \mathrm{cm^{-2}}$, slightly above the cross-over point, above which the gate doping becomes effective. Normalized by the length of graphene, our mobility is quite comparable to the reported results from equally-sized graphene devices encapsulated inside hexagonal boron nitride \cite{shalom2016quantum}.

Figure \ref{fig:4} shows our measured graphene conductivity $\sigma$ as a function of charge carrier density on log-log scale. The red and blue curves correspond to the hole and electron transport, respectively. The residual charge density is extracted below the cross-over point at which the conductivity levels off when approaching the Dirac point. The two curves are fitted using the power-law form $\sigma\propto n^{\gamma}$, with the exponent $\gamma$ close to $1/2$. For the hole transport data, the existence of  $pn$ interfaces at the edges results in a $V_g$-dependent series resistance which is difficult to subtract away with good accuracy. Thus, the residual charge density eventually is evaluated from the electron side data as this unipolar regime is less complex. At the cross-over point, an ultra-low residual charge density of $n{_0}\approx 9\times10^8\ \mathrm{cm^{-2}}$ is obtained, which is close to that of exfoliated graphene devices \cite{kumar2018unconventional,du2008approaching,kumaravadivel2016signatures}. Thus, the main issue for high-quality in the suspended geometry is not the source of the graphene, but the successful current annealing.

On top of the smooth varying background conductance, we observe fast-oscillating Fabry-P\'{e}rot-interference-induced modulation of the differential conductance that is displayed in figure \ref{fig:5}a. The red curve displays the differential conductance $G=dI/dV$ at the $n$ doping side where a clear oscillation riding on the increasing conductance curve can be seen. To extract the oscillations, we employ an advanced adaptive algorithm \cite{Huang1998The} which effectively separates the smooth, sloped background $G_{bg}$ and the oscillations $G_{osc}$. The resulting relative oscillation $G_{osc}/G_{bg}$, shown in blue in figure \ref{fig:5}a, originates from Fabry-P\'{e}rot interference within the cavity delineated by strongly modulated doping regions at the edges \cite{Sonin2009Effect,Cheianov2006Selective,Gunlycke2008Graphene}. In our graphene device, sharp $n-n'$ interfaces are formed between the graphene underneath the contact and the suspended region. The observed, well-developed FP oscillations in figure \ref{fig:5}a indicate that the electron transport in our sample takes place ballistically \cite{rickhaus2013ballistic,du2008approaching,oksanen2014single,shalom2016quantum}.

The FP oscillation periods in gate voltage $ \Delta V_g $ can be extracted from the $G_{osc}/G_{bg}$ trace in figure \ref{fig:5}a.  The obtained $ \Delta V_g $ is displayed as a function of $V_g$  by the blue joined circles in figure \ref{fig:5}b. According to reference \cite{Sonin2009Effect,Laitinen2015Klein}, the oscillation period $ \Delta V_g $ obeys
\begin{equation}\label{period}
\Delta V_g=\frac{e}{C_g}\frac{2a}{1+aL\sqrt{\frac{e}{\pi C_g}}\frac{1}{\sqrt{V_g}}},
\end{equation}
where $L=0.75$ $\mathrm{ \mu m}$ is the length of the graphene stripe, $C_g=1.3\times10^{-5}\ \mathrm{F/m^2}$ is the specific capacitance of the gate electrode, the barrier parameter $a$ is governed by the slope of the electrostatic potential in the junction region with spatially varying doping, and $\hbar$ denotes Planck's constant. Trapeziodal potential barrier with height  $V_0$ together with linearly varying edges over distance yields $a=\sqrt{\pi}|V_0|/\zeta_Fd$, where the parameter $\zeta_F=\sqrt{\pi}\hbar\upsilon_F/e\approx1.27\times10^{7} \ \mathrm{Vm}$ is the Fermi electric flux \cite{Sonin2009Effect,Laitinen2015Klein}. A fit of equation \ref{period} to the data  is illustrated by the red curve in figure  \ref{fig:5}b which yields $a=1.4\times 10^{14}\ \mathrm{1/m^2}$. This value for $a$ is within a factor of $4$ from the result of suspended device fabricated using exfoliated graphene on Si/SiO$_2$ substrate \cite{Laitinen2015Klein}, which yields $a\sim 5\times 10^{14}\ \mathrm{1/m^2}$ at $V_g=2-6 \ \mathrm{V}$. Even though there is some discrepancy, the theoretical result (dashed green curve) of equation \ref{period} with $a=5\times 10^{14}\ \mathrm{1/m^2}$ is still within the scatter of our experimental data. The disagreement is presumably due to variation in the trapezoidal barrier caused by a difference in the contact doping due to the chosen Au/Cr/Au ratios of the contact metal structures.

\section{Summary}
\label{Summary}

We have demonstrated a simple fabrication scheme for suspending two-dimensional (2D) materials with a high success rate achieved by combining CVD graphene and a clean dry transfer method on LOR-coated substrates. The shift of the Dirac peak upon current annealing was systematically monitored and the observed behavior was attributed to  hydrogenation of graphene by H$_2^+$ and low-molecular-weight hydrocarbon radicals. Transport measurement results in our devices indicate a low residual charge density of $9\times10^8$ $\mathrm{cm^{-2}}$. Our samples display a mobility of $1.9 \times 10^5 \mathrm{\ cm^2/Vs}$ and clear FP oscillations, which indicates nearly pristine graphene behavior with ballistic electron transport. The proposed fabrication methods hold remarkable potential for the development of technological applications using suspended structures of 2D materials.

\ack{We thank S. Paraoanu and E. Sonin for useful discussions. Our work was supported by the Academy of Finland (Contracts 310086 and 314448), by the National Natural Science Foundation of China (No. 51675528), and by the National Key R\&D Program of China (No. 2016YFF0203400). This work benefitted from the use of the Aalto University Low Temperature Laboratory infrastructure, which is part of the European Microkelvin Platform funded by European Union's Horizon 2020 Research and Innovation Programme under Grant Agreement No. 824109.}

\section*{Referencing\label{except}}

\bibliographystyle{iopart-num}
\bibliography{reference}   

\end{document}